\documentclass{article}

\usepackage{multirow}
\usepackage{algorithmic}
\usepackage{algorithm}
\usepackage{amsmath}
\usepackage{array}
\usepackage{varwidth}
\usepackage[hyphens]{url}
\usepackage{hyperref}
\usepackage{caption}
\usepackage{numprint}

\newcommand{\II}{\mathcal{I}}
\def\etal{\textit{et al.}}
\def\ie{\textit{i.e.}}

\newcommand{\defword}[1]{\textbf{\boldmath{#1}}}

\newcommand\WrapMultirowEntry[3]{
  \multirow{#1}*{
    \begin{varwidth}{#2em}
      \flushright #3
\end{varwidth}}}

\begin{document}

\npthousandsep{ }

\title{Measuring the Size of Large \\No-Limit Poker Games}

\author{Michael Johanson}

\maketitle

\begin{abstract}

In the field of computational game theory, games are often compared in terms of their size.  This can be measured in several ways, including the number of unique game states, the number of decision points, and the total number of legal actions over all decision points.  These numbers are either known or estimated for a wide range of classic games such as chess and checkers.  In the stochastic and imperfect information game of poker, these sizes are easily computed in ``limit'' games which restrict the players' available actions, but until now had only been estimated for the more complicated ``no-limit'' variants.  In this paper, we describe a simple algorithm for quickly computing the size of two-player no-limit poker games, provide an implementation of this algorithm, and present for the first time precise counts of the number of game states, information sets, actions and terminal nodes in the no-limit poker games played in the Annual Computer Poker Competition.

\end{abstract}

\section{Introduction}
\label{sec:introduction}

Over the last decade, Texas hold'em poker has become a challenge problem and common testbed for researchers studying artificial intelligence and computational game theory.  Poker has proved popular for this task because it is a canonical example of a game with imperfect information and stochastic outcomes.  Since 2006, the Annual Computer Poker Competition (ACPC)~\cite{acpc,acpc_website} has served as a venue for researchers to play their poker agents against each other, revealing which artificial intelligence techniques are effective in practice.  The competition has driven research in the field of computational game theory, resulting in algorithms capable of finding close approximations to optimal strategies in ever larger games.

The size of a game is a simple heuristic that can be used to describe its complexity and compare it to other games, and a game's size can be measured in several ways.  The most commonly used measurement is to count the number of \defword{game states} in a game: the number of possible sequences of actions by the players or by chance, as viewed by a third party that observes all of the players' actions.  In the poker setting, this would include all of the ways that the players private and public cards can be dealt and all of the possible betting sequences.  This number allows us to compare a game against other games such as chess or backgammon, which have $10^{47}$ and $10^{20}$ distinct game states respectively (not including transpositions)\cite{wiki:gamecomplexity}.

In imperfect information games, an alternate measure is to count the number of decision points, which are more formally called more formally called \defword{information sets}.  When a player cannot observe some of the actions or chance events in a game, such as in poker when the opponent's private cards are unknown, many game states will appear identical to the player.  Each such set of indistinguishable game states forms one information set, and an agent's strategy or policy for a game must necessarily depend on its information set and not on the game state: it cannot choose to base its actions on information it does not know.  State-of-the-art algorithms for approximating optimal strategies in imperfect information games, such as Counterfactual Regret Minimization (\defword{CFR})\cite{cprg:cfr}, converge at a rate that depends on the total number of information sets.  

An additional measure related to the number of information sets is the number of legal actions summed across each of the information sets, which we will refer to as the number of \defword{infoset-actions}.  This measure has practical implications on the memory required to store or compute a strategy.  An agent's strategy can be represented as a \defword{behavioral strategy} by storing a probability of taking each legal action at each information set.  Approximating an optimal strategy using a standard CFR implementation requires two double-precision floating point variables per infoset-action: one to store the accumulated regret, and the other to store the average strategy\footnote{Some recent CFR variants, such as CFR-BR~\cite{cprg:cfrbr}, or Oskari Tammelin's PureCFR which uses integers instead of double-precision floats, may require less memory.}.  

In some poker variants it is simple to compute the number of game states and information sets in the game, and counting the number of infoset-actions is not much harder.  For example, in \defword{limit} poker games such as heads-up limit Texas hold'em, the number of information sets can be easily calculated with the single closed-form expression, as we will describe further in Section~\ref{sec:limit}.  This calculation is straightforward because the possible betting actions and information sets within one round are independent of the betting history on previous rounds, and so an expression to calculate the number of game states can be stated for each round as the product of the possible chance events, the number of betting sequences to reach the round, and the number of information sets within the round.  In the ACPC's heads-up limit Texas hold'em events, this can be performed by hand to measure the size of the game at $3.162 \times 10^{17}$ game states and $3.194 \times 10^{14}$ information sets.  In practice, researchers use a lossless state-space abstraction technique that merges states with isomorphic cards, leading to a strategically equivalent but smaller game with $1.380 \times 10^{13}$ information sets and $3.589 \times 10^{13}$ infoset-actions.

In no-limit poker variants, however, measuring the size of the game has until now been computationally challenging.  In these games, the players are provided with a fixed amount of money (a \defword{stack size}) at the start of each game, and may make any number of betting actions of almost any size during any round until they have committed their entire stack.  This means that the possible betting sequences cannot be neatly decomposed by round as is possible in limit poker games.  Since 2007, the ACPC has played three different no-limit poker games, each of which was (correctly) presumed to be far larger than the limit Texas hold'em variants.  The variant played in 2007 and 2008, \$1-\$2 no-limit Texas hold'em with \$1000 (500-blind) stacks, was previously estimated by Gilpin~\etal~to have $10^{71}$ game states~\cite{cmu:nolimit}.  However, the exact size of this game, or of the 2009 and 2010-Present games, has not previously been computed.

In this technical report, we will present for the first time an algorithm that can be used to count the number of game states, information sets, and infoset-actions in these large two-player no-limit poker games.  The algorithm is simple to implement, and source code will be provided along with this technical report.  In Section~\ref{sec:limit}, for context we will briefly describe how the size of heads-up limit poker games are computed.  In Section~\ref{sec:algorithm} we describe the new algorithm, which uses dynamic programming to avoid traversing the game tree.  In Section~\ref{sec:nolimit} we will use our implementation to compute for the first time the exact counts of the game states, information sets, and infoset-actions in the 2007, 2008-2009 and 2010-Present ACPC heads-up no-limit poker games.  Finally, we will briefly discuss the ongoing challenges for action abstraction research in this domain, and propose a new no-limit game as a convenient research testbed for future work.

\section{Measuring heads-up limit games}
\label{sec:limit}

Over the last decade, heads-up limit Texas hold'em has become a common testbed for researchers studying computational game theory in imperfect information games, with significant efforts towards approximating optimal strategies for the game~\cite{cprg:psopti,cprg:cfr,cprg:cfrbr,cmu:egt1}.  In the first paper on approximating a Nash equilibrium strategy for the game, Billings~\etal~presented a figure illustrating the branching factor of the game~\cite[Figure 1]{cprg:psopti}.  In this section, we will describe how the size of the game (in game states, information sets, and infoset-actions) can be precisely computed, to give context to our discussion of no-limit poker.

The heads-up limit Texas hold'em game played in the ACPC is a two player game with four rounds and at most four bets per round.  In the first round, the players' \defword{small blind} and \defword{big blind} (an \defword{ante} required to start the game), counts as a bet, and at most three additional bets are allowed.  The public and private cards are dealt out as as normal for Texas hold'em games.  The ACPC uses the Doyle's game convention, in which each player's stack is reset at the start of each game, and their total winnings are accumulated over all of the games.  In the limit poker events, each player's stack is set to be sufficiently large that the maximum number of bets can be made on each round, making the stack size irrelevant for computing the size of the game.

To start our discussion of the size of the game, we present Table~\ref{table:texas_cards} which lists the number of possible ways to deal the private and public cards on each round.  The Total Two-Player column describes the number of ways to deal the private and public cards to both players on each round: $\binom{52}{2} \times \binom{50}{2}$ on the first round, $\binom{52}{2} \times \binom{50}{2} \times \binom{48}{3}$ on the second round, and so on.  The Total One-Player column describes the number of ways to deal the cards from one player's point of view, when the opponent's cards are unknown: $\binom{52}{2}$ on the first round, $\binom{52}{2} \times \binom{50}{3}$ on the second round, and so on.  Finally, the Canonical One-Player column lists the number of \defword{canonical} card combinations from one player's point of view, after losslessly merging \defword{isomorphic} card combinations that are strategically identical.  

\begin{table}
  {\footnotesize
  \begin{tabular}{|c|r|r|r|}
    \hline
    Round   & Total Two-Player      & Total One-Player    & Canonical One-Player  \\
    \hline
    Preflop & 1,624,350             & 1,326               & 169                   \\ 
    Flop    & 28,094,757,600        & 25,989,600          & 1,286,792               \\ 
    Turn    & 1,264,264,092,000     & 1,221,511,200       & 55,190,538              \\ 
    River   & 55,627,620,048,000    & 56,189,515,200      & 2,428,287,420            \\ 
    \hline
  \end{tabular}
  }
  \caption{Possible public and private card combinations in Texas hold'em poker games.}
  \label{table:texas_cards}
\end{table}

Next, we note that in poker games, the betting actions available to the players are independent of the cards that they have been dealt.  This means that the possible action sequences on each round can be enumerated on their own, and then multiplied by the number of card combinations to find the number of game states.  Further, since the players start with a large enough stack that the maximum number of bets can be made on each round, this means that the possible betting sequences within one round are independent of the actions made by the players on earlier rounds.  In Table~\ref{table:limit_sequences}, we present the decision points, terminal nodes, and action sequences that continue to the next round in heads-up limit Texas hold'em.  In the Decision Points column, ``-'' represents the first decision in the round, and ``c'' and ``r'' respectively represent the check/call and bet/raise actions by the players that lead to a decision.  The Terminal column lists the betting sequences that end the game in the current round, and the Continuing column lists the betting sequences that continue to the next round.  Note that we do not allow players to fold when not facing a bet, as this is dominated by checking or calling.

\begin{table}
  {\footnotesize
  \begin{tabular}{|c|l|l|l|l|}
    \hline
    Round
      & Sequences
      & Actions
      & Continuing
      & Terminal \\
    \hline
    \multirow{6}*{Preflop}           
       & \WrapMultirowEntry{6}{8}{8: \_, c, cr, crr, crrr, r, rr, rrr} 
       & \WrapMultirowEntry{6}{8}{21: -f, -c, -r, c-c, c-r, cr-f, cr-c, cr-r, crr-f, crr-c, crr-r, crrr-f, crrr-c, r-f, r-c, r-r, rr-f, rr-c, rr-r, rrr-f, rrr-c} 
       & \WrapMultirowEntry{6}{8}{7: cc, crc, crrc, crrrc, rc, rrc, rrrc}  
       & \WrapMultirowEntry{6}{8}{7: f, rf, rrf, rrrf, crf, crrf, crrrf} \\
       & & & & \\
       & & & & \\
       & & & & \\
       & & & & \\
       & & & & \\
    \hline
    \multirow{9}*{Flop, Turn} 
       & \WrapMultirowEntry{9}{8}{10: \_, c, cr, crr, crrr, crrrr, r, rr, rrr, rrrr} 
       & \WrapMultirowEntry{9}{8}{26: -c, -r, c-c, c-r, cr-f, cr-c, cr-r, crr-f, crr-c, crr-r, crrr-f, crrr-c, crrr-r, crrrr-f, crrrr-c, r-f, r-c, r-r, rr-f, rr-c, rr-r, rrr-f, rrr-c, rrr-r, rrrr-f, rrrr-c} 
       & \WrapMultirowEntry{9}{8}{9: cc, crc, crrc, crrrc, crrrrc, rc, rrc, rrrc, rrrrc} 
       & \WrapMultirowEntry{9}{8}{8: rf, rrf, rrrf, rrrrf, crf, crrf, crrrf, crrrrf} \\
       & & & & \\
       & & & & \\
       & & & & \\
       & & & & \\
       & & & & \\
       & & & & \\
       & & & & \\
       & & & & \\
    \hline
    \multirow{8}*{River} 
       & \WrapMultirowEntry{9}{8}{10: \_, c, cr, crr, crrr, crrrr, r, rr, rrr, rrrr} 
       & \WrapMultirowEntry{9}{8}{26: -c, -r, c-c, c-r, cr-f, cr-c, cr-r, crr-f, crr-c, crr-r, crrr-f, crrr-c, crrr-r, crrrr-f, crrrr-c, r-f, r-c, r-r, rr-f, rr-c, rr-r, rrr-f, rrr-c, rrr-r, rrrr-f, rrrr-c} 
       & \WrapMultirowEntry{9}{8}{9: cc, crc, crrc, crrrc, crrrrc, rc, rrc, rrrc, rrrrc} 
       & \WrapMultirowEntry{9}{8}{17: cc, rc, rf, rrc, rrf, rrrc, rrrf, rrrrc, rrrrf, crc, crf, crrc, crrf, crrrc, crrrf, crrrrc, crrrrf} \\
       & & & & \\
       & & & & \\
       & & & & \\
       & & & & \\
       & & & & \\
       & & & & \\
       & & & & \\
       & & & & \\
    \hline
  \end{tabular}
  }
  \caption{Betting sequences in limit hold'em poker games.}
  \label{table:limit_sequences}
\end{table}

The figures in Tables~\ref{table:texas_cards} and~\ref{table:limit_sequences} can be multiplied together to compute the number of game states, information sets, and infoset-actions.  This is done one round at a time, by taking the number of betting sequences and multiplying it by the branching factor due to the chance events.  If we multiply by the number of two-player chance events we obtain the number of game states, while multiplying by the number of one-player chance events results in the number of information sets.  An example of this calculation is shown in Equation~\ref{eqn:texas_limit}, in which we calculate the total number of information sets, $|\II|$, in heads-up limit Texas hold'em poker.  

\begin{align}
  |\II| & = \binom{52}{2} \times 8 \nonumber \\
                    & + \binom{52}{2}\binom{50}{3} \times 7 \times 10 \nonumber \\
                    & + \binom{52}{2}\binom{50}{3}\binom{47}{1} \times 7 \times 9 \times 10 \nonumber \\
                    & + \binom{52}{2}\binom{50}{3}\binom{47}{1}\binom{46}{1} \times 7 \times 9 \times 9 \times 10 \nonumber \\
                    & = 319,365,922,522,608
  \label{eqn:texas_limit}
\end{align}

Similar calculations can be performed to compute the number of game states or the number of infoset-actions, which are presented in Table~\ref{table:texas_limit}.  Of particular interest are the total number of canonical information sets and canonical infoset-actions, as these figures describe the complexity in time and memory of computing an optimal strategy for the game using CFR.  In theory, CFR's convergence bound is linear in the number of canonical information sets~\cite[Theorem 4]{cprg:cfr}.  In practice, a standard CFR implementation requires two double-precision floating point variable per infoset-action: one to accumulate regret, and the other to accumulate the average strategy.  

\begin{table}
  {\footnotesize
  \begin{tabular}{|c|c|r|r|r|r|}
    \hline
    \WrapMultirowEntry{6}{5}{Betting Sequences}   & Round   & Sequences         & Sequence-Actions  & Continuing        & Terminal \\
    & Preflop & 8                 & 21                & 7                 & 7        \\
    & Flop    & 70                & 182               & 63                & 56       \\
    & Turn    & 630               & 1638              & 567               & 504      \\
    & River   & 5670              & 14742             & 0                 & 9639     \\
    & Total   & 6378              & 16583             &                   & 10206    \\
    \hline
    \WrapMultirowEntry{6}{5}{One-Player Canonical} & Round   & Infosets          & Infoset-Actions   & Continuing        & Terminal \\
    & Preflop & 1352              & 3549              & 1183              & 1183     \\
    & Flop    & 9.008e7           & 2.342e8           & 8.107e7           & 7.206e7 \\
    & Turn    & 3.477e10          & 9.040e10          & 3.129e10          & 2.781e10\\
    & River   & 1.377e13          & 3.580e13          & 0                 & 2.341e13\\
    & Total   & 1.380e13          & 3.589e13          &                   & 2.343e13\\
    \hline
    \WrapMultirowEntry{6}{5}{One-Sided}            & Round   & Infosets          & Infoset-Actions   & Continuing        & Terminal \\
    & Preflop & 10608             & 27846             & 9282              & 9282     \\
    & Flop    & 1.819e9           & 4.730e9           & 1.637e9           & 1.455e9\\
    & Turn    & 7.696e11          & 2.001e12          & 6.926e11          & 6.156e11\\
    & River   & 3.186e14          & 8.283e14          & 0                 & 5.416e14\\
    & Total   & 3.194e14          & 8.304e14          &                   & 5.422e14\\
    \hline
    \WrapMultirowEntry{6}{5}{Two-Player} & Round   & States            & State-Actions     & Continuing        & Terminal \\
    & Preflop & 1.299e7           & 3.411e7           & 1.137e7           & 1.137e7 \\
    & Flop    & 1.967e12          & 5.113e12          & 1.770e12          & 1.573e12 \\
    & Turn    & 7.965e14          & 2.071e15          & 7.168e14          & 6.372e14\\
    & River   & 3.154e17          & 8.201e17          & 0                 & 5.362e17\\
    & Total   & 3.162e17          & 8.221e17          &                   & 5.368e17\\
    \hline
  \end{tabular}
  }
  \caption{Game size figures for heads-up limit Texas hold'em.}
  \label{table:texas_limit}
\end{table}

The game's size of $3.589 \times 10^{13}$ canonical infoset-actions means that 33 terabytes of disk (using one byte per infoset-action) would be required to store a behavioral strategy, and CFR would require 523 terabytes of RAM (two 8-byte doubles per infoset-action) to solve the game precisely.  While this makes the exact, lossless computation intractable with conventional hardware, it is at least conceivable that such a computation will be possible in time with hardware advances.  Additionally, the size of the game is sufficiently small that unabstracted best response computations have recently become possible~\cite{cprg:rgbr}, and significant progress is being made towards closely approximating an optimal strategy while using state-space abstraction techniques~\cite{cprg:cfrbr}.  

\section{Measuring large no-limit games}
\label{sec:algorithm}

We now turn to the problem of measuring the size of large two-player no-limit poker games.  Unlike the limit poker game discussed in Section~\ref{sec:limit}, no-limit poker presents additional challenges that prevent us from using a single, simple expression as in Equation~\ref{eqn:texas_limit}.  The difficulty is that the possible betting sequences available in each round depend on the betting sequence taken in earlier rounds; furthermore, there can be an enormous number of betting sequences leading to the start of the final round, precluding the approach of simply enumerating them.

The heads-up no-limit poker games played in the ACPC are parameterized by two variables: the stack size that each player has at the start of the game, and the value of the big blind, with the small blind being set equal to half of a big blind.  Each of these variables is measured in dollars, and the stack size is typically a multiple of the big blind.  Unlike in limit Texas hold'em, where each player can only fold, call, or raise a predetermined amount at each decision, no-limit poker allows for a large number of actions.  Each player may fold, call, or bet any whole dollar amount in a range from a \defword{min-bet} to all of their remaining chips.  The size of a min-bet is context-dependent: if a bet has not yet been placed in the current round then a min-bet is defined as equal to the big blind; otherwise, it is equal to the size of the previous bet after calling any outstanding bet.  This means that bets cannot decrease in size during a round.  One exception is that a player is always allowed to bet all of their remaining chips, even if this is smaller than a min-bet.  Once the players have each bet all of their chips (\ie, they are \defword{all-in}), their only legal actions are to call for the remaining rounds until the game is over.  When we present the size of no-limit games, we do not include these trivial information sets or their forced actions.

At any decision point, the actions available to the players depend on the betting history in the game so far: not only on the actions take in the current round, as in limit poker, but on the actions in earlier rounds, as these earlier actions determine the remaining money that the players can use to bet with.  Walking the betting tree of large no-limit games is intractable, as the games are simply far too large.  However, there is still structure to the betting that can be exploited for the purposes of counting the possible states in the game without explicitly walking the tree.  We highlight two critical properties that make this computation possible.  First, a player's legal actions at any decision depend on only three factors: the amount of money they have remaining, the size of the bet that they are facing, and if a check is legal (\ie, if it is the first action in a round).  Within one betting round, any two decision points that are identical in these three factors will have the same legal actions and the same betting subtrees for the remainder of the game, regardless of other aspects of their history.  Second, each of these factors only increases or decreases during a round.  A player's stack size only decreases as they make bets or call an opponent's bets.  The bet being faced is zero at the start of a round (or if the opponent has checked), and can only remain the same or increase during a round.  Finally, a check is only allowed as the first action of a round.  

These observations mean that we do not have to walk the entire game in order to count the decision points.  Instead of considering each betting history independently, we will instead consider the relatively small number of possible configurations of round, stack-size, bet-faced, and check-allowed, and do so one round at a time, starting from the start of the game.  We will incrementally compute the number of action histories that reach each of these configurations by using dynamic programming.  This involves a base case and an inductive step.  The base case is simple: there is one way to reach the start of the game, at which the first player has a full stack minus a small blind, is facing a bet equal to the big blind minus the small blind, and a check is allowed.  Next is the inductive step: if we know that there are $n$ action sequences that reach a given configuration, then for each legal action at that configuration, we can add another $n$ ways to reach the subsequent configurations.  Due to the second property, that each of the round, stack-size, bet-faced and check-allowed factors only increase or decrease, we can update the configurations in a particular order such that applying the inductive step to a configuration only increases the number of ways to reach configurations that we have not yet examined.  For each round in increasing order, we visit all configurations where checks are allowed first, followed by those where a call ends the round.  Within each of these sets, we update configurations in order from largest stacks remaining to smallest.  Within each subset, we update configurations in order from smallest bets faced to largest.  Since all actions taken from a configuration only update the number of ways to reach configurations later in the ordering, only a single traversal is required in order to update all configurations.

When updating each configuration, we can increment counters for each round that track the number of action sequences that lead to a decision by a player and the total number of infoset-actions.  After traversing the set of configurations over all of the rounds, the resulting values can be multiplied by the branching factor due to the chance events for presented earlier in Table~\ref{table:texas_cards} to find the size of each round.  Adding these values across each round produces the overall size of the game in terms of game states, information sets, infoset-actions, and canonical information sets and canonical infoset-actions.

In practice, this algorithm is straightforward to implement and has reasonable memory and time requirements.  The main memory cost is that of allocating one variable to each configuration of stack-size and bet-faced, which can simply be done using a two-dimensional array.  This array can be reused on each round if we also allocate a one-dimensional array indexed by stack size to track the possible ways to reach the next round.  The type of each of these variables should be chosen with caution, as for nontrivial no-limit poker games, they will quickly surpass the maximum value of a 64-bit unsigned integer.  Double-precision floating point variables may be used, but of course result in floating point inaccuracy and cannot provide a precise count.  Instead, an arbitrary precision integer library can be used so that each variable stores a precise integer count.  In our results and in the implementation accompanying this technical report, we used the GNU Multiple Precision Arithmetic Library (GMP)~\cite{gmp} for this purpose.  

The final consideration of the algorithm is its space and time complexity.  As described above, we need only to store a single variable for each of a relatively small number of configurations.  To compute the size of the largest ACPC no-limit game, played from 2010 to the present, approximately 400 million variables were required (20000 possible stack sizes times 20000 possible bets faced).  Using double-precision floating point variables requires less than 3 gigabytes of RAM; using the GMP library's mpz\_t variables requires six gigabytes at startup, and additional memory during the computation as some variables increase and have to allocate more memory.  In terms of time, only a single traversal of the configurations is required, which is essentially four nested for() loops over the rounds, stack sizes, bets faced, and (to update each configuration) the legal actions.  Measuring the size of the 2007-2008 and 2009 ACPC no-limit games, described below, took 47 seconds and 32 seconds respectively.  Measuring the significantly larger 2010-Present ACPC game took nearly two days.

We have released an open source (BSD-licensed) implementation of the algorithm to accompany this technical report.  It can be found online at either of the following locations:
\begin{itemize}
  \item \url{http://webdocs.cs.ualberta.ca/~johanson/publications/poker/2013-techreport-nl-size/2013-techreport-nl-size.html}
  \item \url{http://webdocs.cs.ualberta.ca/~games/poker/count_nl_infosets.html}
\end{itemize}

\section{Sizes of no-limit poker games}
\label{sec:nolimit}

Having described the algorithm used to measure the size of the games, we can now present our main result: the size of the three no-limit games played in the ACPC since 2007, in terms of game states, information sets, infoset-actions, and canonical information sets and canonical infoset-actions.  We will briefly describe each game and its size, and also present the amount of memory required to store a behavioral strategy and to compute an optimal strategy using CFR.  For each game, we will present a table listing the count for each round in scientific notation, and the overall sizes as precise integers; if exact counts of intermediate variables are required, the accompanying implementation outputs precise values.

Note that in the tables below, the `Sequences', `Infosets' and `States' columns show the total number of \textit{nontrivial} situations, where the player has more than one legal action.  Namely, it does not count the forced moves after the players are both all-in and must check and call for the remainder of the game as the public cards are dealt.  Likewise, the `Actions' columns do not include these forced actions.

\subsection{2007-2008: \$1-\$2 with \$1000 (500-blind) stacks}

In 2007, the ACPC introduced its first no-limit poker game, which used a small blind and big blind of \$1 and \$2 respectively and \$1000 (500-blind) stacks.  This was intentionally chosen to be a large, ``deep-stack'' game, as humans typically consider 100-blind stacks to be a normal size.  Gilpin~\etal~had previously estimated this game to have $10^{71}$ game states, quite close to its actual size of $7.16 \times 10^{75}$ game states.  Note that the first round alone, without considering any card information, has more action sequences than the full four-round game of heads-up limit Texas hold'em has game states.

\begin{table}[h!]
  {\footnotesize
  \begin{tabular}{|c|c|r|r|r|r|}
    \hline
    \WrapMultirowEntry{6}{5}{Betting Sequences}   & Round   & Sequences          & Actions       & Continuing    & Terminal \\
    & Preflop & 8.54665e31	 & 2.564e32	 & 8.54665e31	 & 8.54665e31	 \\
    & Flop    & 4.66162e44	 & 1.39849e45	 & 4.66162e44	 & 4.66162e44	 \\
    & Turn    & 1.61489e54	 & 4.84467e54	 & 1.61489e54	 & 1.61489e54	 \\
    & River   & 1.28702e62	 & 3.86106e62	 & 0	         & 2.57404e62	 \\
    & Total   & 1.28702e62	 & 3.86106e62	 &               & 2.57404e62	 \\
    \hline
    \WrapMultirowEntry{6}{5}{One-Sided Canonical} & Round   & Infosets           & Actions       & Continuing    & Terminal \\
    & Preflop & 1.44438e34	 & 4.33315e34	 & 1.44438e34	 & 1.44438e34	 \\
    & Flop    & 5.99853e50	 & 1.79956e51	 & 5.99853e50	 & 5.99853e50	 \\
    & Turn    & 8.91266e61	 & 2.6738e62	 & 8.91266e61	 & 8.91266e61	 \\
    & River   & 3.12525e71	 & 9.37575e71	 & 0	         & 6.2505e71	 \\
    & Total   & 3.12525e71	 & 9.37575e71	 &               & 6.2505e71	 \\
    \hline
    \WrapMultirowEntry{6}{5}{One-Sided} & Round   & Infosets           & Actions       & Continuing    & Terminal \\
    & Preflop & 1.13329e35	 & 3.39986e35	 & 1.13329e35	 & 1.13329e35	 \\
    & Flop    & 1.21154e52	 & 3.63461e52	 & 1.21154e52	 & 1.21154e52	 \\
    & Turn    & 1.97261e63	 & 5.91782e63	 & 1.97261e63	 & 1.97261e63	 \\
    & River   & 7.2317e72	 & 2.16951e73	 & 0	         & 1.44634e73	 \\
    & Total   & 7.2317e72	 & 2.16951e73	 &               & 1.44634e73	 \\
    \hline
    \WrapMultirowEntry{6}{5}{Two-Sided} & Round   & States             & Actions       & Continuing    & Terminal \\
    & Preflop & 1.38828e38	 & 4.16483e38	 & 1.38828e38	 & 1.38828e38	 \\
    & Flop    & 1.30967e55	 & 3.92901e55	 & 1.30967e55	 & 1.30967e55	 \\
    & Turn    & 2.04165e66	 & 6.12494e66	 & 2.04165e66	 & 2.04165e66	 \\
    & River   & 7.15938e75	 & 2.14781e76	 & 0	         & 1.43188e76	 \\
    & Total   & 7.15938e75	 & 2.14781e76	 &               & 1.43188e76	 \\
    \hline
  \end{tabular}
  }
  \caption{Information Set and Game State counts for the 2007-2008 ACPC no-limit game, \$1-\$2 No-Limit Texas Hold'em with \$1000 (500-blind) stacks.}
  \label{table:acpc2007}
\end{table}

Precise counts:
\begin{itemize}
  \item Game states: \numprint{7159379256300503000014733539416250494206634292391071646899171132778113414200}
  \item Information Sets: \numprint{7231696218395692677395045408177846358424267196938605536692771479904913016}
  \item Canonical Infoset-Actions: \numprint{937575457443070937268150407671117224976700640913137221641272121424098561}
\end{itemize}

Solving this game using a standard CFR implementation (2 double-precision floats per canonical infoset-action) would require \numprint{12408707859239112772721938772275407031368328229870} ($1.241 \times 10^{49}$) yottabytes of RAM.

\subsection{2009: \$1-\$2 with \$400 (200-blind) stacks}

In 2009, the ACPC switched its no-limit game to a game with a smaller stack size.  This had two effects.  First, it was closer to what humans would consider a deep-stack no-limit game.  Second, reducing the stack size resulted in a significantly smaller game which required slightly less action abstraction.

\begin{table}[h!]
  {\footnotesize
  \begin{tabular}{|c|c|r|r|r|r|}
    \hline
    \WrapMultirowEntry{6}{5}{Betting Sequences}   & Round   & Sequences          & Actions       & Continuing    & Terminal \\
    & Preflop & 2.23569e19	 & 6.70708e19	 & 2.23569e19	 & 2.23569e19	 \\
    & Flop    & 9.91129e26	 & 2.97339e27	 & 9.91129e26	 & 9.91129e26	 \\
    & Turn    & 4.9179e32	 & 1.47537e33	 & 4.9179e32	 & 4.91789e32	 \\
    & River   & 2.47216e37	 & 7.41638e37	 & 0	         & 4.94427e37	 \\
    & Total   & 2.47221e37	 & 7.41652e37	 &               & 4.94432e37	 \\
    \hline
    \WrapMultirowEntry{6}{5}{One-Sided Canonical} & Round   & Infosets           & Actions       & Continuing    & Terminal \\
    & Preflop & 3.77832e21	 & 1.1335e22	 & 3.77832e21	 & 3.77832e21	 \\
    & Flop    & 1.27538e33	 & 3.82613e33	 & 1.27538e33	 & 1.27538e33	 \\
    & Turn    & 2.71422e40	 & 8.14264e40	 & 2.71422e40	 & 2.71421e40	 \\
    & River   & 6.00311e46	 & 1.80091e47	 & 0	         & 1.20061e47	 \\
    & Total   & 6.00311e46	 & 1.80091e47	 &               & 1.20061e47	 \\
    \hline
    \WrapMultirowEntry{6}{5}{One-Sided} & Round   & Infosets           & Actions       & Continuing    & Terminal \\
    & Preflop & 2.96453e22	 & 8.89359e22	 & 2.96453e22	 & 2.96453e22	 \\
    & Flop    & 2.5759e34	 & 7.72771e34	 & 2.5759e34	 & 2.5759e34	 \\
    & Turn    & 6.00727e41	 & 1.80218e42	 & 6.00727e41	 & 6.00726e41	 \\
    & River   & 1.38909e48	 & 4.16723e48	 & 0	         & 2.77816e48	 \\
    & Total   & 1.38909e48	 & 4.16723e48	 &               & 2.77816e48	 \\
    \hline
    \WrapMultirowEntry{6}{5}{Two-Sided} & Round   & States             & Actions       & Continuing    & Terminal \\
    & Preflop & 3.63155e25	 & 1.08946e26	 & 3.63155e25	 & 3.63155e25	 \\
    & Flop    & 2.78455e37	 & 8.35366e37	 & 2.78455e37	 & 2.78455e37	 \\
    & Turn    & 6.21753e44	 & 1.86526e45	 & 6.21753e44	 & 6.21751e44	 \\
    & River   & 1.3752e51	 & 4.12555e51	 & 0	         & 2.75038e51	 \\
    & Total   & 1.3752e51	 & 4.12555e51	 &               & 2.75038e51	 \\
    \hline
  \end{tabular}
  }
  \caption{Information Set and Game State counts for the 2009 ACPC no-limit game, \$1-\$2 No-Limit Texas Hold'em with \$4000 (200-blind) stacks.}
  \label{table:acpc2009}
\end{table}

Precise counts:
\begin{itemize}
  \item Game states: \numprint{1375203442350500983963565602824903351778252845259200}
  \item Information Sets: \numprint{1389094358906842392181537788403345780331801813952}
  \item Canonical Infoset-Actions: \numprint{180091019297791288982204479657796281550065385037}
\end{itemize}

Solving this game using a standard CFR implementation (2 double-precision floats per canonical infoset-action) would require \numprint{2383484794528738021376773} ($2.383 \times 10^{24}$) yottabytes of RAM.

\subsection{2010-Present: \$50-\$100 with \$20000 (200-blind) stacks}

Finally, we move to the large game currently played in the ACPC.  In 2010, the ACPC competitors chose to ``inflate'' the game by increasing the size of the blinds and the stack, while keeping the ratio between the blinds and the stack the same.  Since players can bet any dollar integer amount between a min-bet and their remaining stack, this dramatically increased the size of the game: instead of having at most 500 or 200 betting options, they now had up to 20000.  The resulting game is by far the largest no-limit variant of the three.

\begin{table}[h!]
  {\footnotesize
  \begin{tabular}{|c|c|r|r|r|r|}
    \hline
    \WrapMultirowEntry{6}{5}{Betting Sequences}   & Round   & Sequences          & Actions       & Continuing    & Terminal \\
    & Preflop & 2.05342e95	 & 6.16026e95	 & 2.05342e95	 & 2.05342e95	 \\
    & Flop    & 1.01693e121	 & 3.05079e121	 & 1.01693e121	 & 1.01693e121 \\
    & Turn    & 1.12027e138	 & 3.36081e138	 & 1.12027e138	 & 1.12027e138 \\
    & River   & 1.13459e151	 & 3.40376e151	 & 0	         & 2.26917e151	 \\
    & Total   & 1.13459e151	 & 3.40376e151	 &               & 2.26917e151	 \\
    \hline
    \WrapMultirowEntry{6}{5}{One-Sided Canonical} & Round   & Infosets           & Actions       & Continuing    & Terminal \\
    & Preflop & 3.47028e97	 & 1.04108e98	 & 3.47028e97	 & 3.47028e97	 \\
    & Flop    & 1.30858e127	 & 3.92574e127	 & 1.30858e127	 & 1.30858e127 \\
    & Turn    & 6.18283e145	 & 1.85485e146	 & 6.18283e145	 & 6.18283e145 \\
    & River   & 2.7551e160	 & 8.26531e160	 & 0	         & 5.51021e160	 \\
    & Total   & 2.7551e160	 & 8.26531e160	 &               & 5.51021e160	 \\
    \hline
    \WrapMultirowEntry{6}{5}{One-Sided} & Round   & Infosets           & Actions       & Continuing    & Terminal \\
    & Preflop & 2.72284e98	 & 8.16851e98	 & 2.72284e98	 & 2.72284e98	 \\
    & Flop    & 2.64296e128	 & 7.92889e128	 & 2.64296e128	 & 2.64296e128 \\
    & Turn    & 1.36842e147	 & 4.10527e147	 & 1.36842e147	 & 1.36842e147 \\
    & River   & 6.37519e161	 & 1.91256e162	 & 0	         & 1.27504e162	 \\
    & Total   & 6.37519e161	 & 1.91256e162	 &               & 1.27504e162	 \\
    \hline
    \WrapMultirowEntry{6}{5}{Two-Sided} & Round   & States             & Actions       & Continuing    & Terminal \\
    & Preflop & 3.33547e101	 & 1.00064e102	 & 3.33547e101	 & 3.33547e101 \\
    & Flop    & 2.85704e131	 & 8.57113e131	 & 2.85704e131	 & 2.85704e131 \\
    & Turn    & 1.41632e150	 & 4.24895e150	 & 1.41632e150	 & 1.41632e150 \\
    & River   & 6.31144e164	 & 1.89343e165	 & 0	         & 1.26229e165	 \\
    & Total   & 6.31144e164	 & 1.89343e165	 &               & 1.26229e165	 \\
    \hline
  \end{tabular}
  }
  \caption{Information Set and Game State counts for 2010-Present ACPC no-limit game, \$50-\$100 No-Limit Texas Hold'em with \$20000 (200-blind) stacks.}
  \label{table:acpc2010}
\end{table}

Precise counts:
\begin{itemize}
  \item Game states: \numprint{631143875439997536762421500982349491523134755009560867161754754138543071866492234040692467854187671526019435023155654264055463548134458792123919483147215176128484600}
  \item Information Sets: \numprint{637519066101007550690301496238244324920475418719042634144396116764136550474559674075887513367166011522983983431697050644965107911879207553424525286198175080441144}
  \item Canonical Infoset-Actions: \numprint{82653117189901827068203416669319641326155549963289335994852924537125934134924844970514122385645557438192782454335992412716935898684703899327697523295834972572001}
\end{itemize}

Solving this game using a standard CFR implementation (2 double-precision floats per canonical infoset-action) would require \numprint{1093904897704962796073602182381684993342477620192821835370553460959511144423474321165844409860820294170754032777335927196407795204128259033} ($1.094 \times 10^{138}$) yottabytes of RAM.

\section{Discussion}
\label{sec:discussion}
While heads-up limit is sufficiently small that the suboptimality of strategies can now be evaluated conveniently~\cite{cprg:rgbr} and close approximations to an optimal strategy are becoming possible~\cite{cprg:cfrbr}, the situation in the no-limit ACPC events appears bleak.  Even the smallest of the three no-limit variants is far larger than heads-up limit.  This is simply a reality of the domain: the game is intrinsically far more complex, and presents additional challenges for state-space abstraction research.  In particular, the no-limit games emphasize the critical importance of research into action abstraction and translation techniques, in which the game is simplified by merging clusters of similar betting actions together.  In practice, there is likely to be little benefit to an agent's ability to differentiate a \$101 bet from a \$99 bet out of a \$20,000 stack, as opposed to simply using a \$100 bet for both cases.  

In order to make meaningful and measurable progress on abstraction and translation techniques, it would be useful to have an analogue to our ability in heads-up limit to evaluate a computer agent's suboptimality in the unabstracted game.  Specifically, we would like to find or create a no-limit game which has three properties:

\begin{itemize}
  \item Unabstracted best response computations are tractable and convenient, so that the worst-case performance of strategies with abstracted betting (and possibly unabstracted cards) can be evaluated.  This allows us to evaluate our abstraction and translation techniques in isolation from other factors.
  \item Unabstracted equilibrium computations are tractable and convenient.  This would allow us to compute an optimal strategy for the game, and measure its in-game performance against agents that use betting abstraction.
  \item Strategic elements similar to that of no-limit Texas hold'em.  As much as possible, we would prefer our game to have similar card elements and betting structure to the game played in the competition.  This means that when possible, we would prefer a game with multiple rounds, a full-sized (or at least large) deck, 5-card poker hands, and stack sizes large enough that simple jam/fold techniques are not effective~\cite{jamfold}.  Agents that abstract the actions in a straightforward way (such as fold-call-pot-allin, for example) will ideally be demonstrated to be highly exploitable, so that an improvement can be distinguished with additional research on action abstraction techniques.
\end{itemize}

The first property is a strict requirement: for the game to be useful, we need to be able to precisely evaluate agents in the full, unabstracted game.  The second property would be very convenient: if unabstracted equilibria can be closely approximated, then it allows for the meaningful in-game performance comparisons that we will be forced to use in the full-scale no-limit Texas hold'em domain.  We will likely have to be flexible on the final property.  It likely will not be possible to find a four-round game with a full deck and large stack sizes that remains both tractable and interesting; instead, we will have to simplify the game in some way.  As motivation, we can consider the [2-1], [2-4], and [3-1] parameterized limit hold'em games recently proposed by Johanson~\etal~\cite{cprg:pcs}, in which the number of rounds and maximum number of bets per round, respectively, are varied to produce smaller games.  In the no-limit domain, the equivalent parameterization is a [r-\$s] game, where r is the number of rounds and \$s is the stack size.  

\section{2-\$20 \$1-\$2 no-limit royal hold'em: a testbed game for future abstraction research}
\label{sec:royal}

As a final contribution of this technical report, we would like to propose one such small no-limit game that may have the properties that we desire from a new common research testbed game: [2-\$20] \$1-\$2 no-limit royal hold'em.  Royal hold'em is a variant of Texas hold'em played with a 20-card deck containing only the Ten through Ace of each of four suits.  [2-\$20] refers to a 2-round game, with a \$20 stack.  As in Texas hold'em, preflop begins with each player receiving two private cards, and the flop begins with three public cards.  The size of this game is presented below in Table~\ref{table:royalholdem}.

\begin{table}[h!]
  {\footnotesize
    \begin{tabular}{|c|c|r|r|r|r|}
    \hline
    \WrapMultirowEntry{4}{5}{Betting Sequences}   & Round   & Sequences          & Actions       & Continuing    & Terminal \\
    & Preflop & 1188	 & 3561	         & 1187	 & 1187	 \\
    & Flop    & 19996	 & 57616	 & 0	 & 38807	 \\
    & Total   & 21184	 & 61177	 &       & 39994	 \\
    \hline
    \WrapMultirowEntry{4}{5}{One-Sided Canonical} & Round   & Infosets           & Actions       & Continuing    & Terminal \\
    & Preflop & 29700	         & 89025	 & 29675	 & 29675	 \\
    & Flop    & 1.55169e08	 & 4.471e08	 & 0	         & 3.01142e08	 \\
    & Total   & 1.55199e08	 & 4.47189e08	 &               & 3.01172e08	 \\
    \hline
    \WrapMultirowEntry{4}{5}{One-Sided} & Round   & Infosets           & Actions       & Continuing    & Terminal \\
    & Preflop & 225720	         & 676590	 & 225530	& 225530	 \\
    & Flop    & 3.10018e09	 & 8.93278e09	 & 0	        & 6.01664e09	 \\
    & Total   & 3.10041e09	 & 8.93346e09	 &              & 6.01686e09	 \\
    \hline
    \WrapMultirowEntry{4}{5}{Two-Sided} & Round   & States             & Actions       & Continuing    & Terminal \\
    & Preflop & 3.45352e07	 & 1.03518e08	 & 3.45061e07	 & 3.45061e07	 \\
    & Flop    & 3.25519e11	 & 9.37942e11	 & 0	         & 6.31747e11	 \\
    & Total   & 3.25553e11	 & 9.37942e11	 &               & 6.31781e11	 \\
    \hline
  \end{tabular}
  }
  \caption{Information Set and Game State counts for [2-\$20] \$1-\$2 no-limit royal hold'em.}
  \label{table:royalholdem}
\end{table}

This game is small enough that CFR would only require 7 gigabytes of RAM, making it tractable on consumer-grade computers, and a common testbed domain that can be shared by all ACPC competitors.  While it is tempting to consider larger games that would require 256 gigabytes of RAM to solve, this would make the game intractable to all but the largest academic research groups competing in the ACPC.  The number of game states in this game is significantly smaller than that of heads-up limit Texas hold'em, and so real game best response computations should be no slower and likely will be considerably faster.  It remains to be shown whether or not this game is sufficiently ``interesting'', by which we mean that simple jam-fold strategies and heavily abstracted agents would ideally be both exploitable by a best response and lose to an unabstracted equilibrium.  If simple strategies are effective in the game, then more complex games involving a larger stack size may have to be considered, balanced against the exponentially growing memory requirement.

\section{Conclusion}
\label{sec:conclusion}

Heads-up no-limit Texas hold'em poker has become a significant research domain since the introduction of a no-limit poker event in the Annual Computer Poker Competition in 2007.  However, even the simple measurement of the size of the game in terms of game states, information sets, and actions has proved difficult, and previously could only be estimated.  In this technical report, we presented an algorithm that can efficiently and exactly compute the size of the ACPC no-limit poker games without requiring exhaustive game tree traversals.  We presented the size of the three no-limit poker variants played in the ACPC since 2007, and discussed the need for a small testbed domain that would help motivate state-space abstraction research into these very large domains.

\bibliographystyle{abbrv}
\bibliography{13-nl-size-techreport}

\end{document}